\documentclass[12pt,preprint]{aastex}

\usepackage{natbib}

\usepackage{color}

\received{}
\revised{}
\accepted{}
\shorttitle{ALMA TW Hya Molecular Lines}
\shortauthors{Nomura et al.}

\begin{document}

\title{High Spatial Resolution Observations of Molecular Lines towards the Protoplanetary Disk around TW Hya with ALMA}

\author{Hideko Nomura\altaffilmark{1,2,11},
  Takashi Tsukagoshi\altaffilmark{1,11}
  Ryohei Kawabe\altaffilmark{1,2}, 
Takayuki Muto\altaffilmark{3}, 
Kazuhiro D. Kanagawa\altaffilmark{4},
Yuri Aikawa\altaffilmark{5}, Eiji Akiyama\altaffilmark{6}, 
Satoshi Okuzumi\altaffilmark{7}, Shigeru Ida\altaffilmark{8}, 
Seokho Lee\altaffilmark{1}, Catherine Walsh\altaffilmark{9} and
T.J. Millar\altaffilmark{10}}

\email{hideko.nomura@nao.ac.jp}










\altaffiltext{1}{National Astronomical Observatory of Japan, 2-21-1 Osawa, Mitaka, Tokyo 181-8588, Japan}
\altaffiltext{2}{Department of Astronomical Science, The Graduate University for Advanced Studies, SOKENDAI, 2-21-1 Osawa, Mitaka, Tokyo 181-8588, Japan}
\altaffiltext{3}{Division of Liberal Arts, Kogakuin University, 1-24-2 Nishi-Shinjuku, Shinjuku-ku, Tokyo, 163-8677, Japan}
\altaffiltext{4}{College of Science, Ibaraki University, 2-1-1 Bunkyo, Mito, Ibaraki 310-8512, Japan}
\altaffiltext{5}{Department of Astronomy, Graduate School of Science, The University of Tokyo, Bunkyo, Tokyo 113-0033, Japan}
\altaffiltext{6}{Division of Fundamental Education and Liberal Arts, Department of Engineering, Niigata Institute of Technology, 1719 Fujihashi, Kashiwazaki, Niigata, 945-1195, Japan}
\altaffiltext{7}{Department of Earth and Planetary Sciences, Tokyo Institute of Technology, 2-12-1 Ookayama, Meguro, Tokyo, 152-8551, Japan}
\altaffiltext{8}{Earth-Life Science Institute, Tokyo Institute of Technology, 2-12-1 Ookayama, Meguro, Tokyo 152-8550, Japan}
\altaffiltext{9}{School of Physics and Astronomy, University of Leeds, Leeds LS2 9JT, UK}
\altaffiltext{10}{Astrophysics Research Centre, School of Mathematics
and Physics, Queen's University Belfast, University Road, Belfast BT7 1NN, UK}
\altaffiltext{11}{These authors contributed equally to this work.}



\begin{abstract}
  We present molecular line observations of $^{13}$CO and C$^{18}$O $J=3-2$, CN $N = 3 - 2$, and CS $J=7-6$ lines in the protoplanetary disk around TW Hya at a high spatial resolution of $\sim 9$ au (angular resolution of 0.15''), using the Atacama Large Millimeter/Submillimeter Array. 
  A possible gas gap is found in the deprojected radial intensity profile of the integrated C$^{18}$O line around a disk radius of $\sim 58$ au, slightly beyond the location of the au-scale dust clump at $\sim 52$ au, which resembles predictions from hydrodynamic simulations of planet-disk interaction. In addition, we construct models for the physical and chemical structure of the TW Hya disk, taking account of the dust surface density profile obtained from high spatial resolution dust continuum observations. As a result, the observed flat radial profile of the CN line intensities is reproduced due to a high dust-to-gas surface density ratio inside $\sim 20$ au. Meanwhile, the CO isotopologue line intensities trace high temperature gas and increase rapidly inside a disk radius of $\sim 30$ au. A model with either CO gas depletion or depletion of gas-phase oxygen elemental abundance is required to reproduce the relatively weak CO isotopologue line intensities observed in the outer disk, consistent with previous atomic and molecular line observations towards the TW Hya disk. {Further observations of line emission of carbon-bearing species, such as atomic carbon and HCN, with high spatial resolution would help to better constrain the distribution of elemental carbon abundance in the disk gas.}
\end{abstract}



\section{Introduction} \label{sec:intro}

Protoplanetary disks are the birth place of planets. Atacama Large Millimeter/Submillimeter Array (ALMA) observations with high spatial resolution are now revealing the detailed physical and chemical structure of these disks. The TW Hya system is an ideal target to study the planet formation environment due to its proximity (59.5 pc). This disk is well studied with various observations: ALMA observations of dust continuum emission at high spatial resolution have revealed the ring and gap structures in the disk, which could be caused by planet formation \citep{Andrews16, Tsukagoshi16}. In addition, an au-scale dust clump, resembling a circumplanetary disk, was discovered at the distance of 52 au from the central star \citep{Tsukagoshi19}. On the other hand, various molecular and atomic lines, including HD lines, have been observed towards the disk from UV to millimeter wavelengths, and suggest the depletion of the carbon and oxygen elemental abundances in the gas-phase \citep[e.g.,][]{Bergin13, Du15, Kama16, Zhang19, Calahan21, Lee21}. The gas-phase elemental abundances of protoplanetary disks are the key to understand the composition of atmospheres of gas giants formed in the disk midplane \citep[e.g.,][]{Madhusudhan19, Notsu20}.

In this work we present ALMA observations of molecular lines towards the TW Hya disk with an angular resolution of 0.15'' corresponding to a spatial resolution of $\sim 9$ au at the distance of TW Hya,  {the second highest spatial resolution molecular line observations towards protoplanetary disks, the CO $J=3-2$ observations of TW Hya being the highest with the beam size of {0.139''$\times$0.131''} \citep{Huang18}}, thanks to its proximity. The observations aim to search for the gas substructures associated with the substructures seen in dust continuum emission. In addition, we performed model calculations to investigate the effect of the radial distribution of dust grains and also the effect of the elemental abundances of carbon and oxygen in the disk on the molecular abundance distributions and resultant molecular line intensity profiles. 





\section{Observations and Data Reduction}

TW Hya was observed with ALMA in Band 7 on 2016 October 5, December
5 and 6 in Cycle 4 with array configurations of C40-6 and C40-3,
respectively (2016.1.01495.S), and 2018 October 14 in Cycle 6 with
an array configuration of C43-6 (2018.1.01644.S). The total on-source
times were 52, 43, and 45 min.,  {and the baseline coverages were $15-704$m, $18-3143$m, and $37-2516$m,} respectively.
The spectral windows were centered at 329.295 GHz (SPW1), 330.552 GHz
(SPW0), 340.211 GHz (SPW2), and 342.846 GHz (SPW3), covering
C$^{18}$O $J = 3 - 2$, $^{13}$CO $J = 3 - 2$, CN $N = 3 - 2$, and CS
$J = 7 - 6$, respectively. The channel spacing was
$\delta\nu=122.07$ kHz and the bandwidth was 234.38 MHz, except
for SPW3, in which a channel spacing of $\delta\nu=244.141$ kHz
and a bandwidth of 468.75 MHz were used. J1058+0133 was observed
as a bandpass calibrator, while J1037-2934
was used for phase and gain calibration. The mean flux density
of J1037-2934 was $0.67\pm 0.07$ Jy during the observation period.

The visibility data were reduced and calibrated using CASA,
versions 5.4, 5.5, and 5.7. 
%
  {Before concatenating all the calibrated visibilities, three measurement sets were reduced separately.
The data reduction was done using the script provided by ALMA except for the C40-6 data of 2016.1.01495.S in which the calibration script was created by using the analysisUtils package.}
The visibility data were separately reduced for each SPW, and the continuum visibilities were extracted by averaging the line-free channels in all SPWs.
  {After the data reduction, we made the CLEAN map of continuum emission of each measurement set, and the peak position of the emission was measured by fitting with a 2D gaussian function. Then, the measurement sets were concatenated after the field centers were corrected.}

  {For the concatenated data, we first made the CLEAN map of the continuum emission with Briggs weighting with a robust parameter of 0.5.
We employed the multiscale clean with scales of 0 (point source), 0.12 and 0.36 arcsec. We manually created the CLEAN masks, and the CLEAN process was stopped at approximately $2-3 \sigma$ noise level measured at the emission-free region.
We applied an iterative self-calibration to improve the image sensitivity.
First, the visibility phase was solved iteratively by changing solution intervals from 600 to 120s, and finally the amplitude was solved.}

  {The result of the self-calibration was applied to the line data.
The channel maps of the lines were reconstructed with the same imaging parameters as those of the continuum map (Briggs weighting, robust = 0.5, down to 2-3 $\sigma$).
After making maps, all channel maps were smoothed to be at 0.15" resolution. The moment maps were made by integrating with a $3 \sigma$ mask (see below for the velocity ranges of the integration).}


\section{Observed Radial Profiles of Molecular Line Intensities}

The maps of the $^{13}$CO and C$^{18}$O $J=3-2$, CN $N = 3 - 2$,  {and CS $J=7-6$} lines are obtained with the synthesized beam size of 0.15''$\times$0.15''.
The 1$\sigma$ RMS noise levels in 0.12 km s$^{-1}$ width-channels are 2.2 mJy beam$^{-1}$ ($^{13}$CO), 2.8 mJy beam$^{-1}$ (C$^{18}$O), 1.9 mJy beam$^{-1}$ (CN) and 1.7 mJy beam$^{-1}$ (CS), respectively. The peak brightness temperature map of each line was obtained using the data without the dust continuum subtraction. In addition, the integrated intensity maps were created by integrating from 0.60 to 5.04 km s$^{-1}$ ($^{13}$CO), from 1.32 to 4.56 km s$^{-1}$ (C$^{18}$O), from 1.20 to 5.28 km s$^{-1}$ (CN),
and from 1.20 to 4.08 km s$^{-1}$ (CS), using the dust continuum subtracted data. The resultant noise levels of the maps were 2.3 mJy beam$^{-1}$ km s$^{-1}$ ($^{13}$CO), 2.4 mJy beam$^{-1}$ km s$^{-1}$ (C$^{18}$O), and 2.3 mJy beam$^{-1}$ km s$^{-1}$ (CN), and 2.0 mJy beam$^{-1}$ km s$^{-1}$ (CS).  {The integrated intensity map (moment 0 map) of each line with dust continuum subtraction is plotted in Fig.~1.}

The deprojected radial profiles of (a) the brightness temperature at the line peaks and (b) the integrated line intensities for the $^{13}$CO $J=3-2$, C$^{18}$O $J=3-2$, CN $N=3-2$ and CS $J=7-6$ lines are plotted in Fig.~2. The $^{13}$CO and C$^{18}$O lines are strong near the central star, while the CN line has a relatively flat distribution up to $> 80$ au \citep[see also][]{Teague20}. The line distributions are consistent with our previous observations that had more than two times lower spatial resolution \citep{Nomura16}. With our higher spatial resolution, the enhancement of the intensities of the $^{13}$CO and C$^{18}$O lines near the central star are emphasized, and the peak brightness temperatures have increased from 40K to 65K, and from 30K to 40K for the $^{13}$CO and C$^{18}$O lines, respectively. The peak brightness temperatures of the CO isotopologue lines decrease inside $\sim 10$ au, similar to the behavior of the $^{12}$CO $J=3-2$ line \citep{Huang18}. This could be due to the line broadening in the inner disk where the gradient of the Keplerian rotation velocity within the beam size is large, which results in the decrease of the peak brightness temperature. In the outer disk, the CN line intensity is stronger by more than an order of magnitude and around a factor of five, compared with those of the $^{13}$CO line and the C$^{18}$O line, respectively.

The dust continuum observations at high spatial resolution ($< 0.1''$) have revealed gap and ring structures, but no clear substructure near the dust gaps and rings is found in the molecular line observations probably due to relatively lower spatial resolution ($\sim 0.15''$). Meanwhile, the C$^{18}$O line intensity profile has a possible substructure of a gap around a disk radius of 58 au (Fig.~2c, see the discussion in Sect.~5.1). The CS line intensity also has a gap around 60 au, while there is no gap in the $^{13}$CO and CN line intensities in this region. This is probably because the $^{13}$CO line is optically thick and does not trace the column density profile. Also, the high peak brightness temperature of the CN line suggests that it traces the disk surface and that it is possibly optically thick (see also Sect.~4).

{We note that the 3 $\sigma$ mask adopted for the moment maps affects the results especially when the dust continuum subtracted lines are weak. For example, the CS line intensity inside a disk radius of $\sim 10$ au becomes $3-4$ times higher {if a mask is not used.} The C$^{18}$O and CS line intensities in the outer disk also become higher by $10-20$\%.}


\section{TW Hya Disk Model}

We model the physical and chemical structure of the TW Hya disk to study which models are preferable to reproduce the observed flat radial profile of the CN line intensity, the enhancement of the CO isotopologue line intensities in the inner disk, as well as the high intensity ratios of the CN line to the CO isotopologue lines in the outer disk.

\subsection{Physical Model}

A self-consistent model of the density and temperature profiles of gas and dust is constructed for an axisymmetric Keplerian disk around a T Tauri star with $M_*=0.5M_{\odot}$, $R_*=2R_{\odot}$, and $T_*=4000$K.  {The dust temperature and the gas temperature profiles are obtained by assuming local radiative equilibrium and local thermal balance, respectively, taking into account the irradiation from the central star. The gas density profile is calculated self-consistently with the gas temperature profile under the assumption of hydrostatic equilibrium in the vertical direction in the disk.} The stellar FUV and X-ray models are based on the observations towards TW Hya \citep[see][]{Nomura05, Nomura07}. The gas surface density profile is obtained by assuming a steady viscous accretion disk with a mass accretion rate of $\dot{M}=10^{-8}M_{\odot}$yr$^{-1}$ and a viscous parameter of $\alpha=0.01$. The outer edge is assumed to follow an exponential tail. The total gas mass is $1.3\times 10^{-2} M_{\odot}$ inside a disk radius of 100 au. For the dust surface density, we consider two models: (D1) the classical model that adopts the uniform gas-to-dust mass ratio of 100, and (D2) the TW Hya model that is based on the observations of dust continuum emission with high spatial resolution \citep{Tsukagoshi16}. Fig.~3 shows the assumed dust surface density profiles.  {The TW Hya model (D2) is within a factor of 2 of the dust surface density profile obtained from the observations (shaded region in light blue) for which the 190 GHz dust continuum data in \cite{Tsukagoshi16} and the 336 GHz dust continuum data of this observation are used inside and outside, respectively, a disk radius of 40 au. We simply assume a dust temperature profile of $T(R)=22{\rm K}(R/10 {\rm au})^{-0.4}$, where $R$ is the disk radius, and a dust opacity of 1.3 cm$^2$ g$^{-1}$ and 4.0 cm$^2$ g$^{-1}$ at 190 GHz and 336 GHz, respectively, referring to the opacity of the $a_{\rm max}=1$mm dust model in \cite{Draine06}, in order to derive the dust surface density profile.} The TW Hya model (D2) is very different from that of the classical model (D1) and the dust grains are concentrated in the inner disk.
Nonetheless, we show results for the model (D1) for reference to understand how the different dust surface density profiles affect the molecular abundance distributions and the line observations.
For the dust size distribution model, we simply assume the MRN model,
$dn/da\propto a^{-3.5}$ ($a$ is the dust radius) \citep{Mathis77}
with a maximum dust radius of $a_{max}=1$mm. Also, the gas and dust are assumed to be well-mixed, that is, the gas-to-dust mass ratio is uniform in the vertical direction for both models (D1) and (D2). 
 {We note that the well-mixing of large dust grains and gas is a simplified assumption, but since the chemistry is mainly affected by the amount of small dust grains, which dominates the total surface area of dust grains \citep[e.g., ][]{Aikawa06}, this assumption is not expected to significantly affect the resulting molecular abundance distributions. However, it could affect the resulting dust temperature distribution, and thus the line intensity profiles.}


The obtained two dimensional distributions of the gas temperature and density, and the FUV radiation field for the different dust surface density profiles (D1) and (D2) are plotted in Fig.~4. The differences between (D1) and (D2) appear mainly beyond $\sim 20$ au. For the TW Hya model (D2), the dust grains are strongly concentrated inside $\sim 20$ au, and also the dust surface density is higher than the classical model (D1) inside $\sim 70$ au, which attenuates the FUV radiation field. As a result, the high gas temperature region is shifted upward beyond $\sim 20$ au in the TW Hya model (D2), and also it affects the gas density distribution. 


\subsection{Chemical Model}

Based on these physical models, molecular abundance distributions are calculated using a detailed chemical reaction network. The gas-phase reaction rates are taken from the UMIST Database for Astrochemistry ``Rate06'' \citep{Woodall07, Walsh10} and surface reactions are not included. However, the freeze-out of gas-phase species onto dust grains as well as the thermal and non-thermal desorption of icy molecules into gas-phase are taken into account. As the initial condition for the calculations of the chemical reactions, atomic and ionized species are adopted \citep[see Table 8 of][]{Woodall07}.
 {The elemental abundance of sulfur is modified so that the results of the model calculations better fit the observed CS line intensity (see Section 4.3).}
There are many observations of molecular and atomic lines, including HD lines, towards TW Hya, which suggest that the carbon and oxygen elemental abundances are depleted in the gas-phase.
Therefore, we consider three kinds of models for the carbon and oxygen elemental abundances: (A1) no depletion (O/H$=1.76\times 10^{-4}$, C/H$=7.30\times 10^{-5}$), (A2) oxygen depletion (O/H$=8.8\times 10^{-6}$, C/H$=7.30\times 10^{-5}$), and (A3) CO depletion. For the models (A1), the elemental abundances are simply assumed to be uniform all over the disk. For the model (A2), the oxygen elemental abundance is set to be depleted only beyond $\sim 25$ au (the CO snowline is located $\sim 30$ au in this model) in order to reproduce the observations \citep[see also][]{Du15, Bergin16}. For the model (A3), the gas-phase carbon and oxygen abundances are artificially reduced to 5\% of the undepleted values (O/H$=8.80\times 10^{-6}$, C/H$=3.65\times 10^{-6}$) only inside the CO depletion zone ( {which is highlighted by the white dashed line in Fig.~5g}). 
We adopt this distribution motivated by model calculations which include grain surface chemistry and which show that CO gas is depleted even in regions inside of and upward of the CO snowline \citep[see e.g.,][]{Aikawa15, Furuya14, Schwarz18, Schwarz19, Bosman18}.  {The location of the CO depletion zone is set so that the calculated CO isotopologue line intensities fit the observations.} The time dependent differential equations for the chemical reaction network are solved and results are presented at a time of $10^6$ yr.

Fig.~5 shows the resulting two-dimensional distributions of the gas-phase abundance profiles of CO and CN for the different dust surface density profiles (D1) and (D2) with the no depletion model (A1). Those for the oxygen depletion model (A2) and the CO depletion model (A3) with the TW Hya dust surface density profile (D2) are also plotted. The molecular abundance distributions are significantly affected by the dust surface density profiles. In the TW Hya model (D2), the molecular abundance distributions are affected by shadowing beyond a disk radius of $\sim 20$ au, and also by the high dust surface density up to $\sim 70$ au. The boundaries between atomic and molecular species shift upwards since the dust grains absorb the FUV radiation which otherwise destroys molecules. Meanwhile, outside $\sim 70$ au, the FUV radiation can penetrate deeper in the disk and the molecular abundant regions extend closer to the disk midplane in the model (D2).
In the oxygen depletion model (A2), the carbon-containing molecules, such as CN, become more abundant as the carbon-to-oxygen elemental abundance ratio is high (C$/$O$>1$) \citep[e.g.,][]{Wei19}.


\subsection{Molecular Lines}

Making use of the obtained density, temperature and molecular abundance distributions, we calculate the line radiative transfer under the LTE assumption. We adopt the RATRAN code \citep{Hogerheijde00} to a disk in Keplerian rotation \citep[e.g.,][]{Notsu16} and use molecular data from the Leiden Atomic and Molecular Database \citep{Schoier05}. A disk inclination angle of 7 degrees is assumed. The radial profiles of the resulting integrated intensity of the $^{13}$CO $J=3-2$, C$^{18}$O  $J=3-2$, and CN $N=3-2$ for the dust surface density profile model (D1) and (D2) with the abundance model (A1) are plotted in Fig.~6(a)(b). Isotope ratios of $^{12}$CO/$^{13}$CO$=67$ and $^{12}$CO/C$^{18}$O$=444$ are assumed \citep{Qi11}. The intensity of the CN $N=3-2$ inside $\sim 20$ au drops in the TW Hya model (D2), consistent with the observations, because the dust-to-gas surface density ratio inside $\sim 20$ au is high in this model. Since CN is abundant in the region where the FUV radiation is strong, that is, where the dust extinction is smaller than a certain value, the CN column density becomes smaller if the dust-to-gas surface density ratio becomes higher, that is, there is not sufficient gas in regions where the dust extinction becomes significant. 
Meanwhile, the CO column density is less sensitive to the FUV radiation field. 
The CO isotopologue lines become optically thick in the high temperature regions inside $\sim 30$ au (above $\sim 3$ mJy beam$^{-1}$ km s$^{-1}$) and the line intensities increase inward.
  {Dust continuum emission is strong inside $\sim 30$ au for the model (D2), and all the molecular lines are embedded in strong dust continuum, which artificially reduces the line intensities after the dust continuum subtraction \citep[e.g.,][]{Weaver18, Notsu19}.}

Beyond $\sim 20$ au, the $^{13}$CO and C$^{18}$O line intensities for the no depletion model (A1) are high compared with the observations. 
Thus, in order to reproduce the observations, the oxygen depletion model (A2) or the CO depletion model (A3) is preferable, consistent with the results of previous atomic and molecular line observations towards TW Hya. The CN line intensity is not affected in model (A3) since it is emitted from the disk surface, while the CN line intensity becomes high in model (A2) due to high abundance of CN (carbon-containing species) (see Sect.~4.2). The difference between the models (A2) and (A3) is within a factor of $\sim 2$, which is consistent with the result in \cite{Cazzoletti18}.

We compare our model results with observations in Fig.~7. All the calculated line intensities are consistent with the observations (black lines) within a factor of 1.5 (shaded region in light blue) for the TW Hya dust model (D2) with  oxygen depletion (A2) or CO depletion model (A3). Meanwhile, the classical dust model (D1) and the TW Hya dust model (D2) with no depletion model (A1) differs from the observations in the $^{13}$CO and C$^{18}$O line intensities beyond the disk radius of $\sim 20$ au. Also, the classical dust model (D1) differs from the observations in the CN line intensity inside $\sim 20$ au. We note that the result of the CN line intensity of the TW Hya dust model (D2) with no depletion model (A1) is almost identical with CO depletion model (A3).
{The dependences of the CO gas depletion on the physical properties of the disk, such as temperature, density, cosmic-ray ionization rates, gas-to-dust mass ratio, and dust size distributions, have been investigated using model calculations \citep[e.g.,][]{Aikawa15, Furuya14, Schwarz18, Schwarz19, Bosman18}. The temperature range within the CO gas depletion zone in our model is higher (up to $\sim 45$ K) compared with that shown in \cite{Bosman18}. \cite{Aikawa15} and \cite{Furuya14} show the dependence of the CO gas depletion zone on the grain sizes, and the CO gas depletion in high temperature regions may suggest that the mass ratio of small dust grains to gas is relatively high in the TW Hya disk (up to the disk radius of $\sim 50$~au).}

For the CS line, we only compare the TW Hya dust model (D2) with the models (A2) and (A3) since we need to modify the sulfur elemental abundance to the values (A2) $2.0\times 10^{-9}$ and (A3) $4.0\times 10^{-9}$, respectively, in order to better fit the observation. Elemental abundance of sulfur in gas-phase is an unknown factor even in molecular clouds, and freeze-out in ice and/or transformation into refractory material have been suggested \citep[e.g., ][]{Navarro20}. Further observations of other sulfur-bearing species are needed in order to better constrain the sulfur chemistry in the disk \citep[e.g., ][]{Semenov18, LeGal19}.

\section{Discussion}

\subsection{Gas Substructure in the Outer Disk}

The location of the possible gas gap (58 au, Fig.~2c) is just beyond the region where the au-scale dust clump was discovered (52 au) in high spatial resolution dust continuum observations \citep{Tsukagoshi19}. Hydrodynamic simulations of planet-disk interaction suggests that the orbital radius of a planet with fast inward migration is shifted inward compared to the location of the gas gap if the migration time is shorter than the gap-opening time \citep[e.g.,][]{Kanagawa20}.
According to these hydrodynamic simulations, the difference between the locations of the planet and the gas gap depends on the planet-to-stellar mass ratio, the local scale height of the gas disk, the local gas surface density, and the turbulent viscosity. If we assume the possible planet mass is around Neptune mass, as is suggested from the size of the observed dust clump ($\sim 1$ au) \citep{Tsukagoshi19}, and adopt a gas surface density of $\sim 2$ g cm$^{-2}$ and a gas scale height of $\sim 0.08$ at 52 au, which are used in the model in Sect.~4, the observed difference between the locations of the au-scale dust clump and the possible gas gap can be explained by a simulation with a turbulent viscous parameter of $\alpha\sim 10^{-4}-10^{-3}$ \citep{Kanagawa20}. 
  {When turbulent viscosity is small enough, a planet can form a secondary gap at a smaller radii than the planet orbit \citep{Bae17, Dong17, Dong18, Zhang18, Kanagawa20}. \cite{Kanagawa20} show that the separation between the location of the planet and the secondary gap is related to the scale height at the location of the secondary gap. In the TW Hya disk, however, there is no signature of the corresponding secondary gap near the au-scale dust clump; if we assume the 41 au gap observed in dust continuum emission is the secondary gap, the scale height must satisfy $H/R\sim 0.03$ (where $H$ is the scale height and $R$ is the disk radius) at 41 au, which is inconsistent with the disk model in Sect.~4.
  This might suggest that the turbulent viscosity is not weaker than $\alpha \sim 10^{-3}$.
We note that this value is consistent with those measured by the observations of molecular line widths \citep[e.g., ][]{Flaherty18}.}
Further investigations are needed for more accurate constraint.

\subsection{Dust Size Distributions and Turbulent Mixing}

The observed low spectral index of dust continuum emission indicates that the dust continuum emission becomes optically thick and/or that large dust grains dominate the opacity in the inner region of the disk ($\leq 20$ au) \citep{Tsukagoshi16, Huang18, Macias21}. Meanwhile, comparison between our model calculations and the observed radial profiles of molecular line intensities suggests that the decrease of the CN line intensity in the inner disk as well as high brightness temperatures of the $^{13}$CO and the CN lines can be explained by the existence of a certain amount of small dust grains at least in the disk surface, which is consistent with the observations of the infrared dust scattering \citep[e.g.,][]{Akiyama15, Rapson15, vanBoekel17}. These results indicate segregation of the dust size distribution in the vertical direction of the disk as suggested by the dust settling model in a turbulent disk \citep[e.g.,][]{Dubrulle95}. The strength of turbulence could also affect the origin of the C$/$O depletion in the disk: the CO gas depletion model, due to chemical reactions and gas-grain interactions, suggests that turbulent mixing will reduce the effect of the CO gas depletion \citep{Furuya14}  {since the ice molecules on grains are photodissociated/photodesorbed when small grains are stirred up into the disk surface. We note that this result depends on the chemical processes in the reaction network and the size of dust grains, and, for example, \cite{Krijt20} show that the CO gas depletion is enhanced in a turbulent disk.}
Meanwhile, the oxygen depletion model, due to water freeze-out on dust grains and the subsequent radial drift of the grains, will become more efficient with turbulent mixing \citep[e.g.,][]{Du15, Bergin16}.

\subsection{Model Calculations of Other Molecular and Atomic Lines}

 {In Fig.~6 (c)(d), we also plot the results of the model calculations of the line intensities of C$_2$H $N=4-3$ and [CI] $^3P_1-^3P_0$ for the oxygen depletion (A2) and the CO depletion (A3) models. These lines were detected towards the TW Hya disk \citep[e.g.,][]{Kama16, Bergin16}. For both models (A2) and (A3) the calculated intensities of the C$_2$H line are consistent with the observed value around a disk radius of $\sim 60$ au, taking into account the difference in the beam sizes of the observations, while the model intensities are stronger than the observed intensity in the inner disk. The calculated intensity of the atomic carbon is about an order of magnitude stronger than the observations for both models (A2) and (A3).
   {Depletion of the gas-phase elemental abundance of carbon in the surface layer of the disk would decrease the calculated intensities of the C$_2$H, and [CI] lines \citep[e.g.,][]{Kama16,Bergin16,Bosman19}. Since the emitting regions of the CN, C$_2$H, and [CI] lines are similar in the models, it is likely that some fine-tuning of the distribution of elemental abundance of carbon is needed in order to fit all the observed line intensities. Further investigations, for example, examining the possibilities of uniform CO depletion in the vertical direction due to turbulent mixing, the adjustment of the gas temperature distribution and/or the dust size and spatial distributions in the disk surface, are needed in order to quantify the elemental abundance of carbon in the disk \citep[e.g.,][]{Trapman21}.}

In addition, we plot the HCN and HCO$^+$ $J=4-3$ line intensities for the oxygen depletion (A2) and the CO gas depletion (A3) models. The relative line intensities of CN to HCN become higher in the model (A2) since CN becomes abundant while HCN is less affected in the condition of C$/$O $> 1$. The difference in the HCO$^+$ line intensity is not very significant between these models.
Further observations, such as relative intensity ratio of CN and HCN lines could help {constrain the carbon-to-oxygen elemental abundance ratio in the disk.}

\section{Summary}

  {We present high spatial resolution observations of $^{13}$CO and C$^{18}$O $J=3-2$, CN $N = 3 - 2$, and CS $J=7-6$ lines in the protoplanetary disk around TW Hya with ALMA. The spatial resolution of $\sim 9$ au (angular resolution of 0.15'') is achieved, and we find a possible gas gap in the deprojected radial intensity profile of the integrated C$^{18}$O line around a disk radius of $\sim 58$ au, just beyond the location of the au-scale dust clump at $\sim 52$ au. This resembles prediction by hydrodynamic simulations of planet-disk interaction, which suggests that the orbital radius of a planet with fast inward migration is shifted inward compared to the location of the gas gap if the migration time is shorter than the gap-opening time. The observed difference between the locations of the au-scale dust clump and the possible gas gap can be explained by the simulation with a turbulent viscous parameter of $10^{-4}-10^{-3}$.}

  {The high resolution observations of molecular lines also find rapid increases of $^{13}$CO and C$^{18}$O line intensities as well as a decrease of CN line intensity in the inner disk. We performed model calculations for the physical and chemical structure of the TW Hya disk, taking account of the dust surface density profile obtained from high spatial resolution dust continuum observations. Our model calculations and the observations of brightness temperatures at the line peaks suggest that the CO isotopologue lines become optically thick and they trace high temperature gas inside a disk radius of $\sim 30$ au. The decrease of the CN line intensity can be interpreted by a high dust-to-gas surface density ratio inside $\sim 20$ au since the attenuation of the FUV radiation by dust grains reduce the CN column density. In the outer disk beyond $\sim 20$ au, we find relatively weak CO isotopologue line intensities compared with the CN line intensity, which can be explained by either CO gas depletion or depletion of gas-phase oxygen elemental abundance. The observations can be reproduced when the CO gas or the oxygen elemental abundance is reduced to 5 \% of those of molecular clouds, which is consistent with the previous atomic and molecular line observations towards the TW Hya disk. The radial profile of the CS line intensity can also be reproduced if we set the gas-phase elemental abundance of sulfur as $(2.0-4.0)\times 10^{-9}$.}

  {Further observations of different molecular and atomic lines with high spatial resolution will reveal more detailed physical and chemical properties of the disk, such as depletion of CO gas and/or gas-phase elemental abundances, dust size distribution and turbulent mixing, as well as the gas substructures around the substructures found in dust continuum emission, and then possible formation process of planets in the disk.}




\acknowledgments

We would like to thank the referee for his/her comments which improved our paper. 
This paper makes use of the following ALMA data: ADS/JAO.ALMA\#2016.1.01495.S and ADS/JAO.ALMA\#2018.1.01644.S.
ALMA is a partnership of ESO (representing its member states), NSF (USA) and NINS (Japan), together with NRC (Canada), NSC and ASIAA (Taiwan) and KASI (Republic of Korea), in cooperation with the Republic of Chile. The Joint ALMA Observatory is operated by ESO, AUI/NRAO and NAOJ.
The numerical calculations in this work were carried out on XC40 at YITP in Kyoto University, and PC cluster at the Center for Computational Astrophysics, National Astronomical Observatory of Japan.
This work is supported by JSPS and MEXT Grants-in-Aid for Scientific Research, 18H05441, 19K03910, 20H00182 (H.N.), 17H01103, 19K03932 (T.M.), 19K14779 (K.K.), and 17K05399 (E.A.), and NAOJ ALMA Scientific Research Grant Number of 2018-10B.
C.W.~acknowledges financial support from the University of Leeds, the 
Science and Technology Facilities Council, and UK Research and 
Innovation (grant numbers ST/R000549/1, ST/T000287/1, and MR/T040726/1).
Astrophysics at QUB is supported by grant ST/P000312/1 from the STFC.

{\it Facility:} \facility{ALMA}.


\begin{thebibliography}

\bibitem[{{Aikawa} {et~al.}(2015){Aikawa}, {Furuya}, {Nomura}, \&
    {Qi}}]{Aikawa15}
{Aikawa}, Y., {Furuya}, K., {Nomura}, H., \& {Qi}, C. 2015, \apj, 807, 120

\bibitem[{{Aikawa \& Nomura}(2006){Aikawa}, \&  {Nomura}}]{Aikawa06}
{Aikawa}, Y. \& {Nomura}, H. 2006, \apj, 642, 1152

\bibitem[{{Akiyama} {et~al.}(2015){Akiyama}, {Muto}, {Kusakabe}, {Kataoka},
  {Hashimoto}, {Tsukagoshi}, {Kwon}, {Kudo}, {Kandori}, {Grady}, {Takami},
  {Janson}, {Kuzuhara}, {Henning}, {Sitko}, {Carson}, {Mayama}, {Currie},
  {Thalmann}, {Wisniewski}, {Momose}, {Ohashi}, {Abe}, {Brandner}, {Brandt},
  {Egner}, {Feldt}, {Goto}, {Guyon}, {Hayano}, {Hayashi}, {Hayashi}, {Hodapp},
  {Ishi}, {Iye}, {Knapp}, {Matsuo}, {Mcelwain}, {Miyama}, {Morino},
  {Moro-Martin}, {Nishimura}, {Pyo}, {Serabyn}, {Suenaga}, {Suto}, {Suzuki},
  {Takahashi}, {Takato}, {Terada}, {Tomono}, {Turner}, {Watanabe}, {Yamada},
  {Takami}, {Usuda}, \& {Tamura}}]{Akiyama15}
{Akiyama}, E., {Muto}, T., {Kusakabe}, N. {et~al.} 2015, \apjl, 802, L17
  
\bibitem[{{Andrews} {et~al.}(2016){Andrews}, {Wilner}, {Zhu}, {Birnstiel},
  {Carpenter}, {P{\'e}rez}, {Bai}, {{\"O}berg}, {Hughes}, {Isella}, \&
  {Ricci}}]{Andrews16}
{Andrews}, S.~M., {et~al.} 2016, \apjl, 820, L40

\bibitem[{{Bae} {et~al.}(2017){Bae}, {Zhu}, \& {Hartmann}}]{Bae17}
{Bae}, J., {Zhu}, Z., \& {Hartmann}, L. 2017, \apj, 850, 201

\bibitem[{{Bergin} {et~al.}(2013){Bergin}, {Cleeves}, {Gorti}, {Zhang},
  {Blake}, {Green}, {Andrews}, {Evans}, {Henning}, {{\"O}berg}, {Pontoppidan},
  {Qi}, {Salyk}, \& {van Dishoeck}}]{Bergin13}
{Bergin}, E.~A., {Cleeves}, L.~I., {Gorti}, U. {et~al.} 2013, \nat, 493, 644

\bibitem[{{Bergin} {et~al.}(2016){Bergin}, {Du}, {Cleeves}, {Blake}, {Schwarz},
  {Visser}, \& {Zhang}}]{Bergin16}
{Bergin}, E.~A., {Du}, F., {Cleeves}, L.~I., {Blake}, G.~A., {Schwarz}, K.,
  {Visser}, R., \& {Zhang}, K. 2016, \apj, 831, 101

\bibitem[{{Bosman} \& {Banzatti}(2019) {Bosman} \& {Banzatti}}]{Bosman19}
  {Bosman}, A. D. \& {Banzatti}, A. 2019, \aap, 632, L10
  
  \bibitem[{{Bosman} {et~al.}(2018){Bosman}, {Walsh}, \& {van Dishoeck}}] {Bosman18}{Bosman}, A. D., {Walsh}, C., \& {van Dishoeck}, E.F. 2018, \aap, 618, A182

\bibitem[{Calahan} {et~al.}(2021){Calahan}, Jenny and {Bergin}, Edwin and {Zhang}, Ke and {Teague}, Richard and {Cleeves}, Ilsedore and {Bergner}, Jennifer and {Blake}, Geoffrey A. and {Cazzoletti}, Paolo and {Guzman}, Viviana and {Hogerheijde}, Michiel R. and {Huang}, Jane and {Kama}, Mihkel and {Loomis}, Ryan and {Oberg}, Karin and {van Dishoeck}, Ewine F. and {Terwisscha van Scheltinga}, Jeroen and {Walsh}, Catherine and {Wilner}, David and {Qi}, Charlie]{Calahan21}
  {Calahan}, J., {et~al.} 2021, \apj, 908, 8

\bibitem[{{Cazzoletti} {et~al.}(2018){Cazzoletti}, {van Dishoeck}, {Visser},
  {Facchini}, \& {Bruderer}}]{Cazzoletti18}
{Cazzoletti}, P., {van Dishoeck}, E.~F., {Visser}, R., {Facchini}, S., \&
  {Bruderer}, S. 2018, \aap, 609, A93

\bibitem[{{Dong} {et~al.}(2017){Dong}, {Li}, {Chiang}, \& {Li}}]{Dong17}
{Dong}, R., {Li}, S., {Chiang}, E., \& {Li}, H. 2017, \apj, 843, 127

\bibitem[{{Dong} {et~al.}(2018){Dong}, {Li}, {Chiang}, \& {Li}}]{Dong18}
---. 2018, \apj, 866, 110

\bibitem[{{Draine}(2006)}]{Draine06}
{Draine}, B.~T. 2006, \apj, 636, 1114

\bibitem[{{Du} {et~al.}(2015){Du}, {Bergin}, \& {Hogerheijde}}]{Du15}
{Du}, F., {Bergin}, E.~A., \& {Hogerheijde}, M.~R. 2015, \apjl, 807, L32

\bibitem[{{Dubrulle} {et~al.}(1995){Dubrulle}, {Morfill}, \&
  {Sterzik}}]{Dubrulle95}
{Dubrulle}, B., {Morfill}, G., \& {Sterzik}, M. 1995, \icarus, 114, 237

\bibitem[{{Flaherty} {et~al.}(2018){Flaherty}, {Hughes}, {Teague}, {Simon},
  {Andrews}, \& {Wilner}}]{Flaherty18}
{Flaherty}, K.~M., {Hughes}, A.~M., {Teague}, R., {Simon}, J.~B., {Andrews},
  S.~M., \& {Wilner}, D.~J. 2018, \apj, 856, 117

\bibitem[{{Furuya} \& {Aikawa}(2014)}]{Furuya14}
{Furuya}, K., \& {Aikawa}, Y. 2014, \apj, 790, 97

\bibitem[{{Hogerheijde} \& {van der Tak}(2000)}]{Hogerheijde00}
{Hogerheijde}, M.~R., \& {van der Tak}, F.~F.~S. 2000, \aap, 362, 697

\bibitem[{{Huang} {et~al.}(2018){Huang}, {Andrews}, {Cleeves}, {{\"O}berg},
  {Wilner}, {Bai}, {Birnstiel}, {Carpenter}, {Hughes}, {Isella}, {P{\'e}rez},
  {Ricci}, \& {Zhu}}]{Huang18}
{Huang}, J., {et~al.} 2018, \apj, 852, 122

\bibitem[{{Kama} {et~al.}(2016){Kama}, {Bruderer}, {van Dishoeck},
  {Hogerheijde}, {Folsom}, {Miotello}, {Fedele}, {Belloche}, {G{\"u}sten}, \&
  {Wyrowski}}]{Kama16}
{Kama}, M., {et~al.} 2016, \aap, 592, A83

\bibitem[{{Kanagawa} {et~al.}(2020){Kanagawa}, {Nomura}, {Tsukagoshi}, {Muto},
  \& {Kawabe}}]{Kanagawa20}
{Kanagawa}, K.~D., {Nomura}, H., {Tsukagoshi}, T., {Muto}, T., \& {Kawabe}, R.
  2020, \apj, 892, 83

\bibitem[{{Krijt} {et~al.}(2020){Krijt}, {Bosman}, {Zhang}, {Schwarz},
  {Ciesla}, \& {Bergin}}]{Krijt20}
{Krijt}, S., {Bosman}, A.~D., {Zhang}, K., {Schwarz}, K.~R., {Ciesla}, F.~J.,
  \& {Bergin}, E.~A. 2020, \apj, 899, 134

\bibitem[{{Le Gal} {et~al.}(2019){Le Gal}, {{\"O}berg}, {Loomis}, {Pegues}, \&
  {Bergner}}]{LeGal19}
{Le Gal}, R., {{\"O}berg}, K.~I., {Loomis}, R.~A., {Pegues}, J., \& {Bergner},
  J.~B. 2019, \apj, 876, 72

\bibitem[{{Lee} {et~al.}(2021){Lee}, {Nomura}, {Furuya}, \& {Lee}}]{Lee21}
{Lee}, S., {Nomura}, H., {Furuya}, K., \& {Lee}, J.-E. 2021, \apj, 908, 82

\bibitem[{{Macias} {et~al.}(2021){Macias}, {Guerra-Alvarado},
  {Carrasco-Gonzalez}, {Ribas}, {Espaillat}, {Huang}, \& {Andrews}}]{Macias21}
{Macias}, E., {Guerra-Alvarado}, O., {Carrasco-Gonzalez}, C., {Ribas}, A.,
{Espaillat}, C.~C., {Huang}, J., \& {Andrews}, S.~M. 2021, \aap, in press
(arXiv:2102.04648)

\bibitem[{{Madhusudhan}(2019)}]{Madhusudhan19}
{Madhusudhan}, N. 2019, \araa, 57, 617

\bibitem[{Mathis} {et~al.}(1977)]{Mathis77}
{Mathis}, J.~S. and {Rumpl}, W. and {Nordsieck}, K.~H. 1977, \apj, 217, 425

\bibitem[{{Navarro-Almaida} {et~al.}(2020){Navarro-Almaida}, {Le Gal},
  {Fuente}, {Rivi{\`e}re-Marichalar}, {Wakelam}, {Cazaux}, {Caselli}, {Laas},
  {Alonso-Albi}, {Loison}, {Gerin}, {Kramer}, {Roueff}, {Bachiller},
  {Commer{\c{c}}on}, {Friesen}, {Garc{\'\i}a-Burillo}, {Goicoechea},
  {Giuliano}, {Jim{\'e}nez-Serra}, {Kirk}, {Lattanzi}, {Malinen}, {Marcelino},
  {Mart{\'\i}n-Dom{\`e}nech}, {Mu{\~n}oz Caro}, {Pineda}, {Tercero},
  {Trevi{\~n}o-Morales}, {Roncero}, {Hacar}, {Tafalla}, \&
  {Ward-Thompson}}]{Navarro20}
{Navarro-Almaida}, D., {et~al.} 2020, \aap, 637, A39

\bibitem[{{Nomura} {et~al.}(2007){Nomura}, {Aikawa}, {Tsujimoto}, {Nakagawa},
  \& {Millar}}]{Nomura07}
{Nomura}, H., {Aikawa}, Y., {Tsujimoto}, M., {Nakagawa}, Y., \& {Millar}, T.~J.
  2007, \apj, 661, 334

\bibitem[{{Nomura} \& {Millar}(2005)}]{Nomura05}
{Nomura}, H., \& {Millar}, T.~J. 2005, \aap, 438, 923

\bibitem[{{Nomura} {et~al.}(2016){Nomura}, {Tsukagoshi}, {Kawabe}, {Ishimoto},
  {Okuzumi}, {Muto}, {Kanagawa}, {Ida}, {Walsh}, {Millar}, \& {Bai}}]{Nomura16}
{Nomura}, H., {et~al.} 2016, \apjl, 819, L7

\bibitem[{{Notsu} {et~al.}(2019){Notsu}, {Akiyama}, {Booth}, {Nomura}, {Walsh}, {Hirota}, {Honda}, {Tsukagoshi}, {Millar}}]{Notsu19}
  {Notsu}, S., {et~al.} 2019, \apj, 875, 96
    
\bibitem[{{Notsu} {et~al.}(2020){Notsu}, {Eistrup}, {Walsh}, \&
  {Nomura}}]{Notsu20}
{Notsu}, S., {Eistrup}, C., {Walsh}, C., \& {Nomura}, H. 2020, \mnras, 499,
  2229

\bibitem[{{Notsu} {et~al.}(2016){Notsu}, {Nomura}, {Ishimoto}, {Walsh},
  {Honda}, {Hirota}, \& {Millar}}]{Notsu16}
{Notsu}, S., {Nomura}, H., {Ishimoto}, D., {Walsh}, C., {Honda}, M., {Hirota},
  T., \& {Millar}, T.~J. 2016, \apj, 827, 113

\bibitem[{{Notsu} {et~al.}(2019){Notsu}, {Akiyama}, {Booth}, {Nomura}, {Walsh},
  {Hirota}, {Honda}, {Tsukagoshi}, \& {Millar}}]{Notsu19}
{Notsu}, S., {et~al.} 2019, \apj, 875, 96

\bibitem[{{Qi} {et~al.}(2011){Qi}, {D'Alessio}, {{\"O}berg}, {Wilner},
  {Hughes}, {Andrews}, \& {Ayala}}]{Qi11}
{Qi}, C., {D'Alessio}, P., {{\"O}berg}, K.~I., {Wilner}, D.~J., {Hughes},
  A.~M., {Andrews}, S.~M., \& {Ayala}, S. 2011, \apj, 740, 84

\bibitem[{{Rapson} {et~al.}(2015){Rapson}, {Kastner}, {Millar-Blanchaer}, \&
    {Dong}}]{Rapson15}
{Rapson}, V.~A., {Kastner}, J.~H., {Millar-Blanchaer}, M.~A., \& {Dong}, R.
  2015, \apjl, 815, L26

\bibitem[{{Sch{\"o}ier} {et~al.}(2005){Sch{\"o}ier}, {van der Tak}, {van
  Dishoeck}, \& {Black}}]{Schoier05}
{Sch{\"o}ier}, F.~L., {van der Tak}, F.~F.~S., {van Dishoeck}, E.~F., \&
  {Black}, J.~H. 2005, \aap, 432, 369

\bibitem[{{Schwarz} {et~al.}(2018){Schwarz}, {Bergin}, {Cleeves}, {Zhang},
  {{\"O}berg}, {Blake}, \& {Anderson}}]{Schwarz18}
{Schwarz}, K.~R., {Bergin}, E.~A., {Cleeves}, L.~I., {Zhang}, K., {{\"O}berg},
  K.~I., {Blake}, G.~A., \& {Anderson}, D. 2018, \apj, 856, 85

\bibitem[{{Schwarz} {et~al.}(2019){Schwarz}, {Bergin}, {Cleeves}, {Zhang},
  {{\"O}berg}, {Blake}, \& {Anderson}}]{Schwarz19}
{Schwarz}, K.~R., {Bergin}, E.~A., {Cleeves}, L.~I., {Zhang}, K., {{\"O}berg},
  K.~I., {Blake}, G.~A., \& {Anderson}, D.~E. 2019, \apj, 877, 131

\bibitem[{{Semenov} {et~al.}(2018){Semenov}, {Favre}, {Fedele}, {Guilloteau},
  {Teague}, {Henning}, {Dutrey}, {Chapillon}, {Hersant}, \&
  {Pi{\'e}tu}}]{Semenov18}
{Semenov}, D., {et~al.} 2018, \aap, 617, A28

\bibitem[{{Teague} \& {Loomis}(2020)}]{Teague20}
{Teague}, R., \& {Loomis}, R. 2020, \apj, 899, 157

\bibitem[{{Trapman} {et~al.}(2021)}]{Trapman21}{Trapman}, L., {Bosman}, A.D., {Rosotti}, G., {Hogerheijde}, M. R., \& {van Dishoeck}, E. F. 2021, \aap, in press (arXiv: 2103.05654)

\bibitem[{{Tsukagoshi} {et~al.}(2016){Tsukagoshi}, {Nomura}, {Muto}, {Kawabe},
  {Ishimoto}, {Kanagawa}, {Okuzumi}, {Ida}, {Walsh}, \&
  {Millar}}]{Tsukagoshi16}
{Tsukagoshi}, T., {et~al.} 2016, \apjl, 829, L35

\bibitem[{{Tsukagoshi} {et~al.}(2019){Tsukagoshi}, {Muto}, {Nomura}, {Kawabe},
  {Kanagawa}, {Okuzumi}, {Ida}, {Walsh}, {Millar}, {Takahashi}, {Hashimoto},
  {Uyama}, \& {Tamura}}]{Tsukagoshi19}
---. 2019, \apjl, 878, L8

\bibitem[{{van Boekel} {et~al.}(2017){van Boekel}, {Henning}, {Menu}, {de
  Boer}, {Langlois}, {M{\"u}ller}, {Avenhaus}, {Boccaletti}, {Schmid},
  {Thalmann}, {Benisty}, {Dominik}, {Ginski}, {Girard}, {Gisler}, {Lobo Gomes},
  {Menard}, {Min}, {Pavlov}, {Pohl}, {Quanz}, {Rabou}, {Roelfsema}, {Sauvage},
  {Teague}, {Wildi}, \& {Zurlo}}]{vanBoekel17}
{van Boekel}, R., {et~al.} 2017, \apj, 837, 132

\bibitem[{{Weaver} {et~al.}(2018){Weaver}, {Isella}, \& {Boehler}}]{Weaver18}
{Weaver}, E., {Isella}, A., \& {Boehler}, Y. 2018, \apj, 853, 113

\bibitem[{{Walsh} {et~al.}(2010){Walsh}, {Millar}, \& {Nomura}}]{Walsh10}
{Walsh}, C., {Millar}, T.~J., \& {Nomura}, H. 2010, \apj, 722, 1607

\bibitem[{{Wei} {et~al.}(2019){Wei}, {Nomura}, {Lee}, {Ip}, {Walsh}, \&
  {Millar}}]{Wei19}
{Wei}, C.-E., {Nomura}, H., {Lee}, J.-E., {Ip}, W.-H., {Walsh}, C., \&
  {Millar}, T.~J. 2019, \apj, 870, 129

\bibitem[{{Woodall} {et~al.}(2007){Woodall}, {Ag{\'u}ndez}, {Markwick-Kemper},
  \& {Millar}}]{Woodall07}
{Woodall}, J., {Ag{\'u}ndez}, M., {Markwick-Kemper}, A.~J., \& {Millar}, T.~J.
  2007, \aap, 466, 1197

\bibitem[{{Zhang} {et~al.}(2019){Zhang}, {Bergin}, {Schwarz}, {Krijt}, \&
  {Ciesla}}]{Zhang19}
{Zhang}, K., {Bergin}, E.~A., {Schwarz}, K., {Krijt}, S., \& {Ciesla}, F. 2019,
  \apj, 883, 98

\bibitem[{{Zhang} {et~al.}(2018){Zhang}, {Zhu}, {Huang}, {Guzm{\'a}n},
  {Andrews}, {Birnstiel}, {Dullemond}, {Carpenter}, {Isella}, {P{\'e}rez},
  {Benisty}, {Wilner}, {Baruteau}, {Bai}, \& {Ricci}}]{Zhang18}
{Zhang}, S., {et~al.} 2018, \apjl, 869, L47

\end{thebibliography}


\begin{figure}
  \begin{center}
\includegraphics[scale=.6]{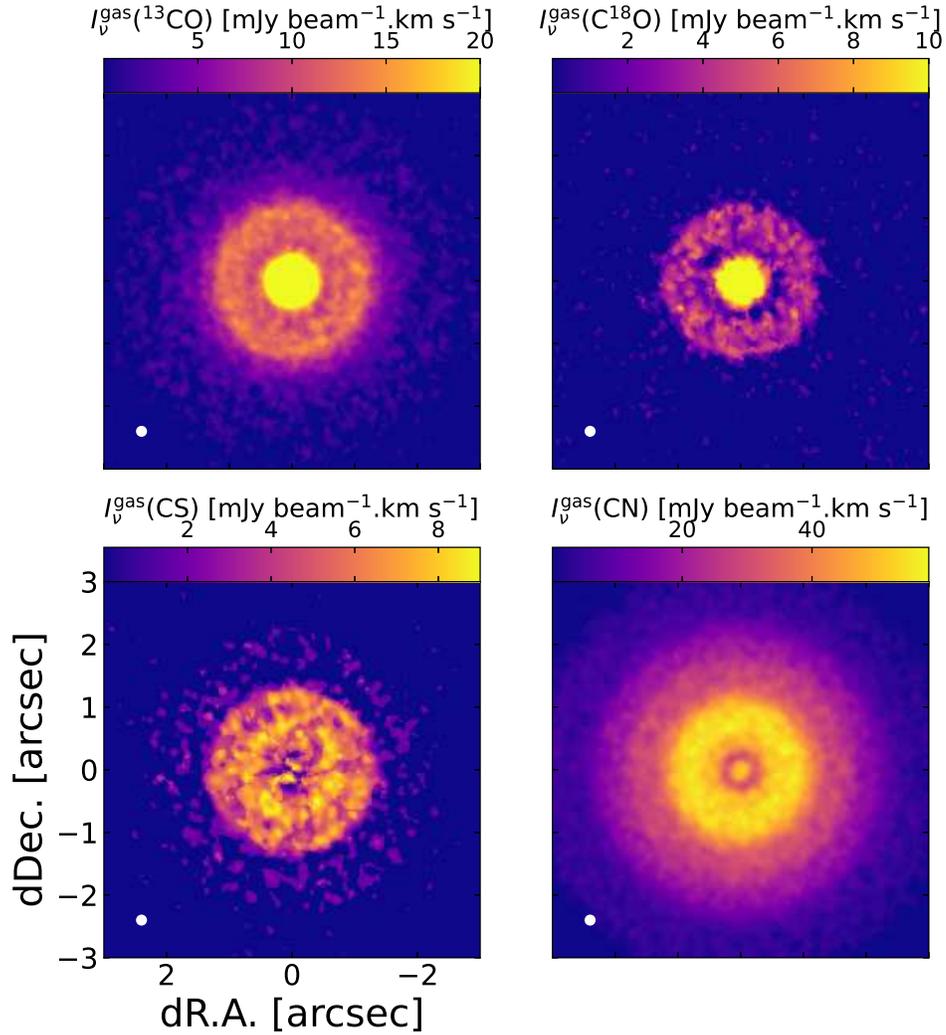}
\caption{ {Integrated intensity maps (moment 0 maps) of (top-left) $^{13}$CO $J=3-2$, (top-right) C$^{18}$O $J=3-2$, (bottom-left) CS $J=7-6$, and (bottom-right) CN $N=3-2$ lines. The synthesized beam size is shown in the bottom left corner of each panel.}
\label{fig1}}
\end{center}
\end{figure}

\begin{figure}
\begin{center}
\includegraphics[scale=.35]{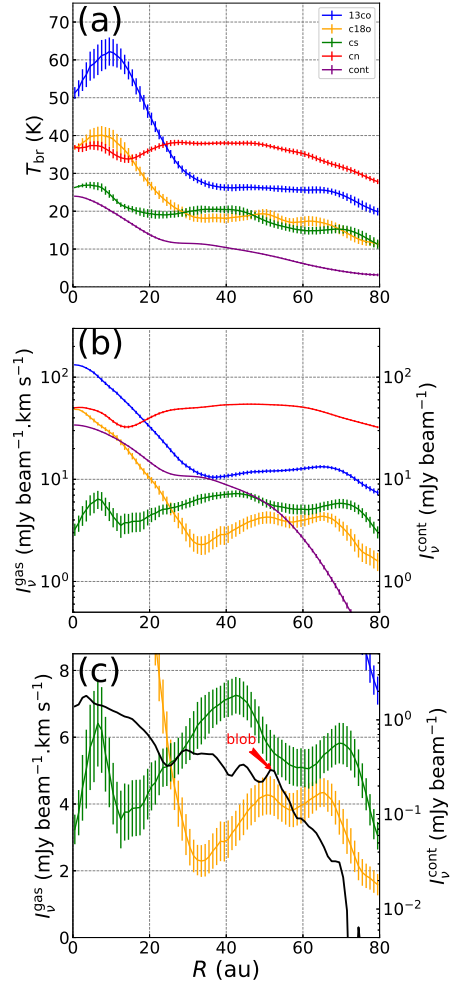}
\caption{Deprojected radial profiles of (a) the brightness temperatures at the line peak and (b) the integrated intensities for the $^{13}$CO $J=3-2$ (blue), C$^{18}$O $J=3-2$ (orange), CN $N=3-2$ (red), and CS $J=7-6$ (green) lines. The brightness temperature and the intensity of dust continuum emission (purple) are also plotted for comparison. The beam size is set as 0.''15$\times$0.''15. (c) Zoom-up of the integrated intensities of the C$^{18}$O (orange) and CS (green) lines around a disk radius of 60 au together with the radial profile of the dust continuum intensity with high spatial resolution (46.88$\times$41.56 mas) along the position angle with the au-scale dust clump (marked with 'blob') (black) \citep[see][]{Tsukagoshi19}. 
\label{fig2}}
\end{center}
\end{figure}


\begin{figure}
\begin{center}
  \hspace*{2cm}\includegraphics[scale=1.0]{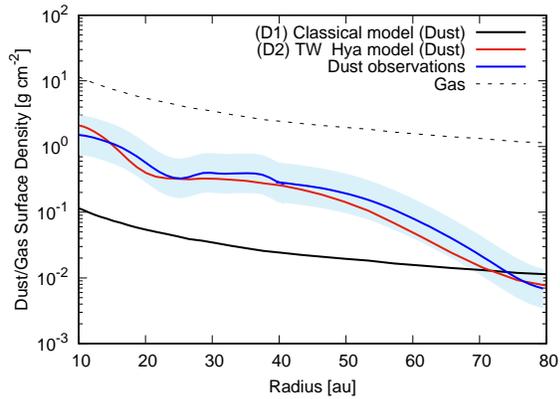}
  \vspace*{3cm}
  \caption{The dust (solid lines) and gas (dashed line) surface density profiles used in the model calculations. The classical model (D1, black) follows the steady-state viscous accretion model and the surface density profile is roughly $\Sigma_{\rm dust}\propto R^{-1}$ (where $R$ is the disk radius). The TW Hya model (D2, red) is based on the high resolution ALMA observations of dust continuum emission \citep{Tsukagoshi16}.  {The dust surface density profile derived from the observations (blue line) is also plotted with a margin for error within a factor of 2 (light blue shaded region).}
\label{fig3}}
\end{center}
\end{figure}


\begin{figure}
\begin{center}
\vspace*{-2cm}
\hspace*{1.5cm}\includegraphics[scale=1.2]{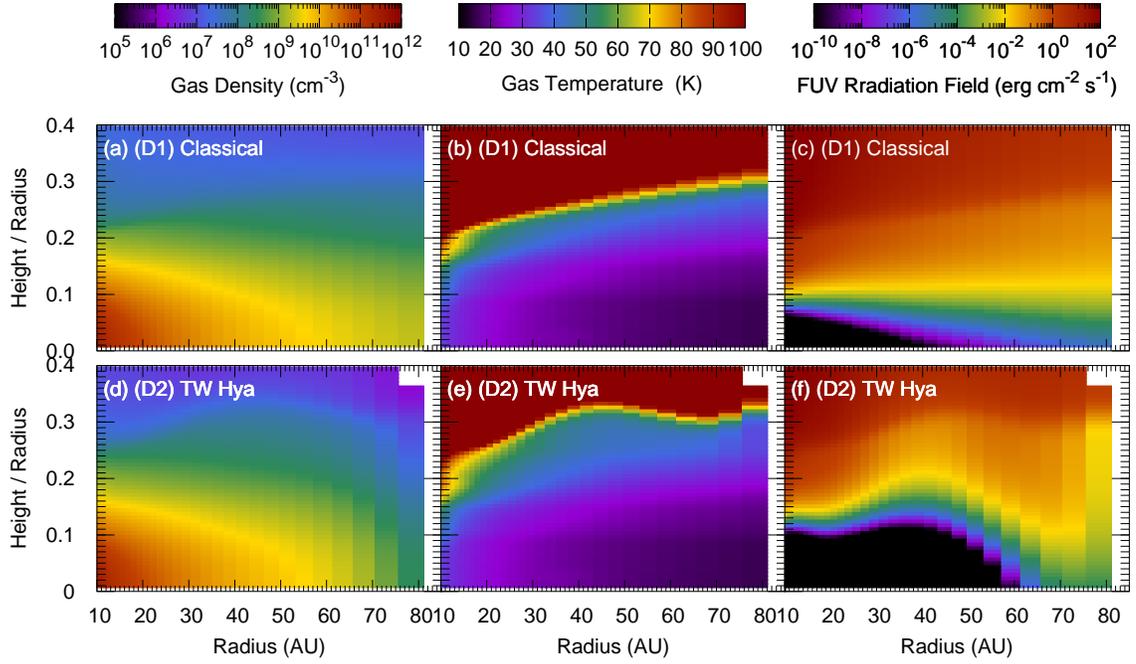}
\vspace*{2cm}
\caption{Two dimensional distributions of (a)(d) the gas density, (b)(e) the gas temperature, and (c)(f) the FUV radiation field for the dust surface density profiles of (D1) the classical model and (D2) the TW Hya model.
\label{fig4}}
\end{center}
\end{figure}


\begin{figure}
\begin{center}
  \includegraphics[scale=1.7]{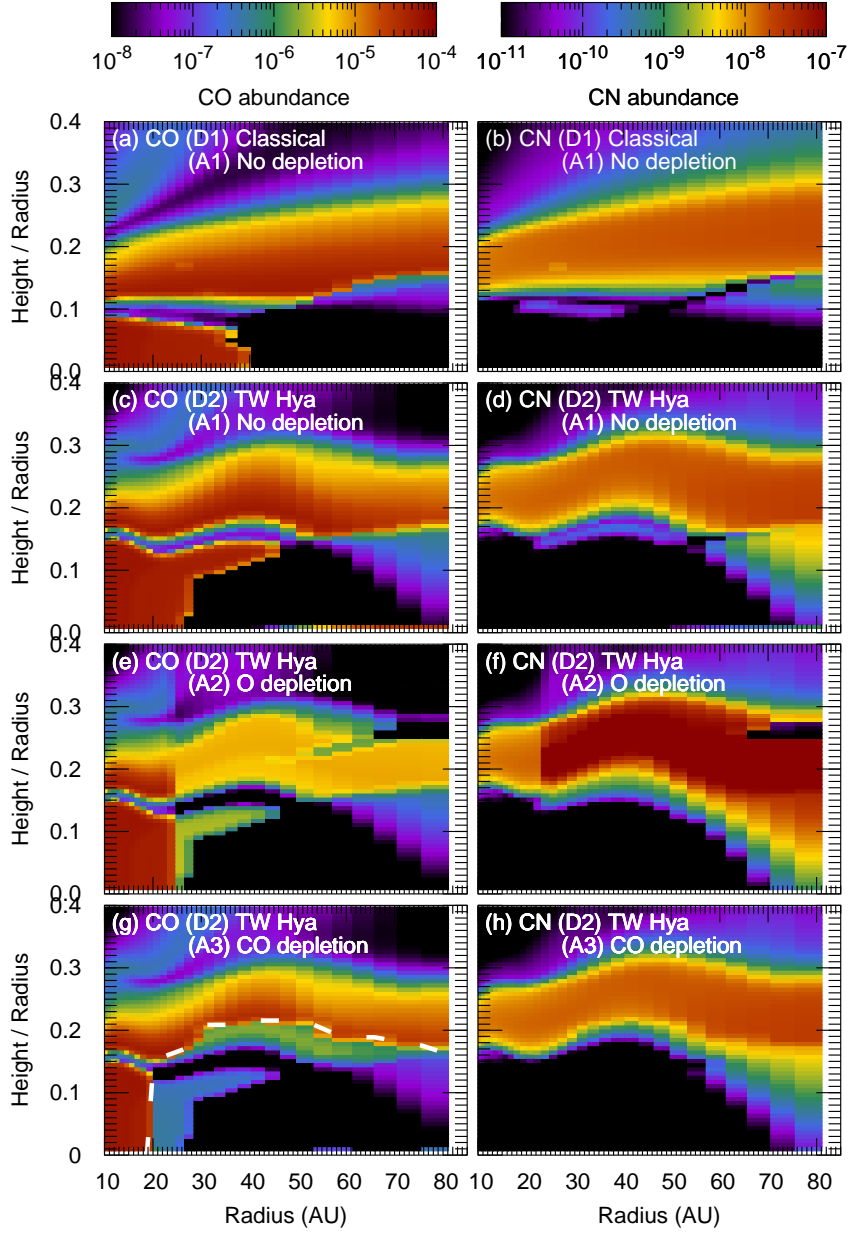}
\caption{Two dimensional distributions of molecular abundances of CO (a),(c),(e),(g) and CN (b),(d),(f),(h). The dust surface density profiles of (D1) the classical model and (D2) the TW Hya model, and the elemental abundance models of (A1) no depletion, (A2) oxygen depletion, and (A3) CO depletion are used. In figure (g) the CO depletion region is plotted with the white dashed line.
\label{fig5}}
\end{center}
\end{figure}


\begin{figure}
\begin{center}
\vspace*{-2cm}
\hspace*{1.5cm}
\includegraphics[scale=1.3]{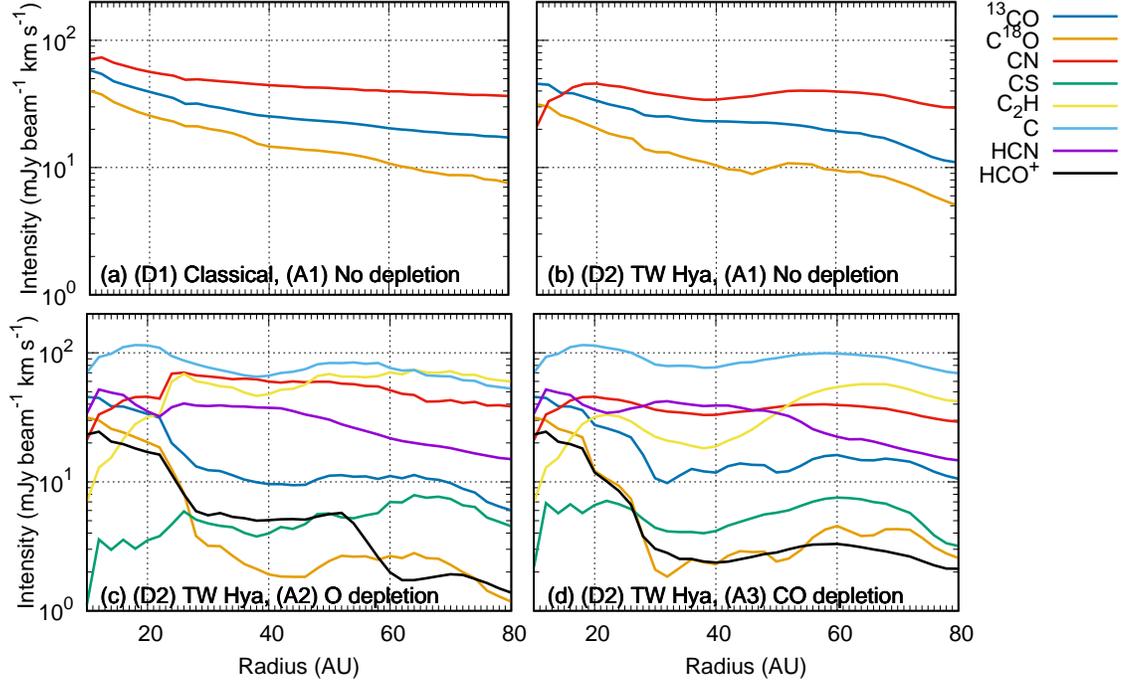}
\vspace*{2.5cm}
\caption{Model calculations of the radial profiles of the integrated intensities for the $^{13}$CO $J=3-2$ (blue), C$^{18}$O $J=3-2$ (orange), CN $N=3-2$ (red),  {and CS $J=7-6$ (green)} lines. The beam size is set as 0.''15$\times$0.''15. The dust surface density profiles of (D1) the classical model and (D2) the TW Hya model, and the elemental abundance models of (A1) no depletion, (A2) oxygen depletion, and (A3) CO depletion are used.
  Model calculations of the integrated intensities for the  {C$_2$H $N=4-3$ (yellow), [CI] $^3P_1-^3P_0$ (light blue),} HCN $J=4-3$ (purple), and HCO$^+$ $J=4-3$ (black) lines are also plotted.
\label{fig6}}
\end{center}
\end{figure}

\begin{figure}
\begin{center}
\vspace*{-2cm}
\hspace*{1.5cm}
\includegraphics[scale=1.3]{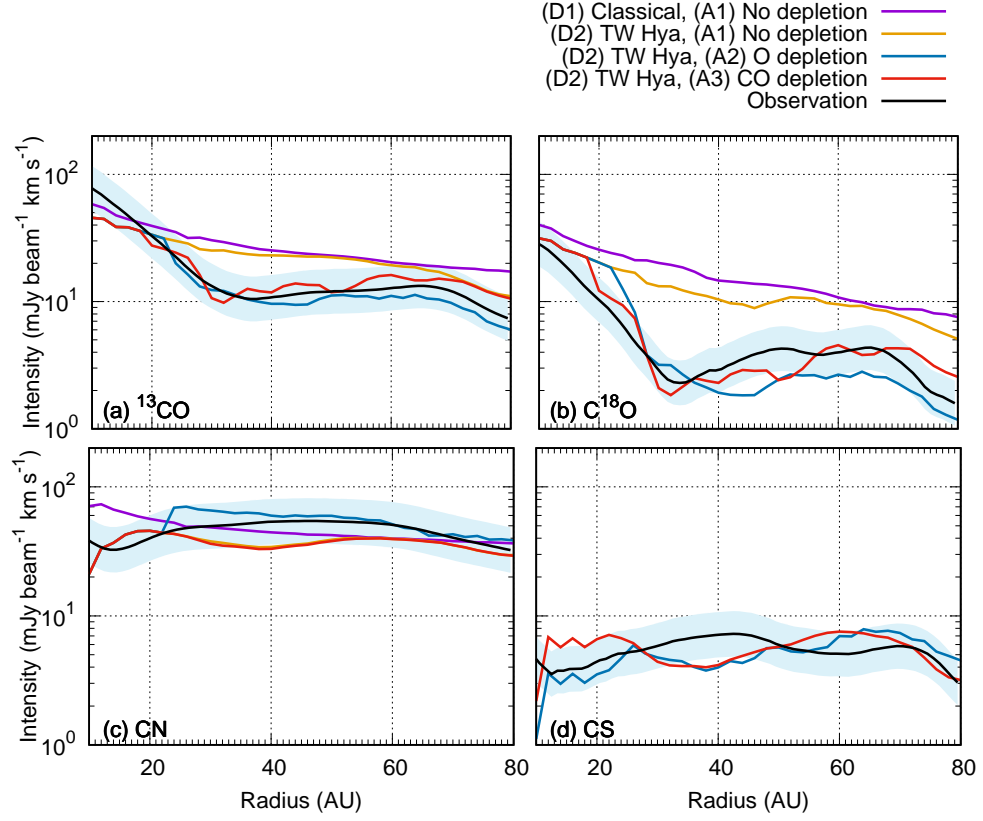}
\vspace*{2.5cm}
\caption{ {Comparison between the model calculations and the observations for the radial profiles of the integrated intensities of (a) $^{13}$CO $J=3-2$, (b) C$^{18}$O $J=3-2$, (c) CN $N=3-2$, and (d) CS $J=7-6$ lines. The dust surface density profiles of (D1) the classical model and (D2) the TW Hya model, and the elemental abundance models of (A1) no depletion, (A2) oxygen depletion, and (A3) CO depletion are used. The light blue shaded regions show a margin for error from the observations within a factor of 1.5.}
\label{fig7}}
\end{center}
\end{figure}

\end{document}